\documentclass[12pt]{iopart}

\usepackage{amsgen}
\usepackage{amssymb}
\usepackage{graphicx}
\usepackage{setstack}
\usepackage{color}

\def\be{\begin{equation}}
\def\ee{\end{equation}}
\def\beq{\begin{equation}}
\def\eeq{\end{equation}}
\def\bea{\begin{eqnarray}}
\def\eea{\end{eqnarray}}

\makeatletter
\@ifundefined{textcolor}{}
{
 \definecolor{BLACK}{gray}{0}
 \definecolor{WHITE}{gray}{1}
 \definecolor{RED}{rgb}{1,0,0}
 \definecolor{GREEN}{rgb}{0,1,0}
 \definecolor{BLUE}{rgb}{0,0,1}
 \definecolor{CYAN}{cmyk}{1,0,0,0}
 \definecolor{MAGENTA}{cmyk}{0,1,0,0}
 \definecolor{YELLOW}{cmyk}{0,0,1,0}
 }

\newcommand{\edit}{}
\newcommand{\td}{{\rm d}} 
\newcommand{\riso}{{\cal{R}}}  \newcommand{\ghn}{{\cal O}}

\newcommand\fscalar[1]{{}^\circ{\! #1}}
\newcommand\fvec[1]{{}^\dagger{\nhalf\nquart #1}} 
\newcommand\ftensor[1]{{}^\ddagger{\nquart #1}}
\newcommand{\negminispace}{\kern-.016667em} 

\newcommand{\half}{\kern.083333em}   
\newcommand{\quart}{\kern.0416675em}  
\newcommand{\nhalf}{\kern-.083333em}   
\newcommand{\nquart}{\kern-.0416675em}  

\newcommand\h{\delta g}
\newcommand\rot{{\Theta}}

\begin{document}

\title{Piecewise Silence in Discrete Cosmological Models}

\author{Timothy Clifton$^1$, Daniele Gregoris$^{2,3,4}$, and Kjell Rosquist$^{2,5}$}
\address{$^1$ School of Physics and Astronomy, Queen Mary University of London, UK.\\
$^2$ Department of Physics, Stockholm University, 106 91 Stockholm, Sweden.\\
$^3$ Max-Planck-Institut f\"{u}r Gravitationsphysik (Albert-Einstein-Institut), Am M\"{u}hlenberg 1, 14476 Potsdam, Germany.\\
$^4$ Erasmus Mundus Joint Doctorate IRAP Ph.D. Student.\\
$^5$ ICRANet, Piazza della Repubblica, 10, I-65122 Pescara, Italy.
}



\begin{abstract}

We consider a family of cosmological models in which all mass is confined to a regular lattice of identical black holes. By exploiting the reflection symmetry about planes that bisect these lattices into identical halves, we are able to consider the evolution of a number of geometrically distinguished surfaces that exist within each of them. {\edit We find that the evolution equations for the reflection symmetric surfaces can be written as a simple set of Friedmann-like equations, with source terms that behave like a set of interacting effective fluids.} We {\edit then} show that gravitational waves are effectively trapped within small chambers for all time, and are not free to propagate throughout the space-time. Each chamber therefore evolves as if it were in isolation from the rest of the universe. We call this phenomenon ``piecewise silence''.

\end{abstract}

\section{Introduction}
\label{sec:int}

Since its birth, classical general relativity has been applied to the construction of cosmological models. This is usually done by specifying the symmetries that are expected to exist on large scales, and then looking for solutions of Einstein's field equations that exhibit those symmetries. The Cosmological Principle, that the Universe is spatially homogeneous and isotropic on large scales, then leads one to the Friedmann-Lema\^itre-Robertson-Walker (FLRW) solutions. Taken as a model for the Universe, the FLRW solutions of Einstein's equations have been shown to be remarkably consistent with a wide array of observations, ranging from the cosmic microwave background \cite{planck}, all the way through to the Hubble diagrams constructed using nearby supernovae \cite{perlmutter,riess}. This consistency, however, is only allowed at the expense of including new and exotic ``dark'' components that must dominate over all other forms of matter in the Universe.

On the other hand, we know that the visible matter in the late Universe is highly inhomogeneous on small scales, being condensed largely into stars that only occupy about $10^{-30}$ of the total volume of space\footnote{Based on a typical stellar density of $\sim 1$g/cm$^{3}$, a baryon fraction of $\sim 0.05$, and a critical density of $\sim 10^{-29}$g/cm$^{3}$.}. The rest is an almost perfect vacuum. This strongly suggests the need to relax our assumptions about homogeneity, at least at some level. Bound up with this is the ``backreaction'' problem, that averaging and evolution do not commute in Einstein's theory \cite{Andersson&Coley:2011}. This means that even if the Universe is close to being homogeneous and isotropic on large scales, there is no guarantee that it will evolve anything like the FLRW solutions of Einstein's equations (unless one adds some otherwise undetectable exotic matter fields, in order to force it to).

The problem that then needs to be addressed, in order to trust the results of the standard approach to cosmology, is to determine the large-scale evolution that emerges in space-times that are close to statistically homogeneous and isotropic on large scales, but very highly inhomogeneous on small scales. This problem has been approached in many different ways in the literature, including using cosmological perturbation theory \cite{perts}, spatial averaging \cite{space}, space-time averaging \cite{space-time}, shortwave approximations \cite{green}, exact solutions \cite{bull}, and approximate solutions \cite{larena}, to name but a few. Our approach is to develop a family of cosmological models that contains a regular array of identical black holes only. These models have the great benefit of admitting a time-symmetric initial value problem that can be solved exactly \cite{Clifton_etal:2012}, as well as allowing for the exact evolution of some high-symmetry curves to be found \cite{Clifton_etal:2013}.

Early work in this spirit includes that of McCrea \cite{mccrea}, Coxeter \& Whitrow \cite{Coxeter&Whitrow:1950} and Lindquist \& Wheeler \cite{Lindquist&Wheeler:1957, geocell}. These latter authors suggested a Wigner-Seitz-type approach in which 3-dimensional space is tiled with regular polyhedra, and a mass is put at the centre of each cell. The geometry of space-time within each cell is then approximated as being Schwarzschild, and approximate junction conditions are used to obtain a large-scale evolution. The optical properties of these models have been studied in \cite{CF1, CF2,CF3}. More recently numerical solutions to Einstein's equations have been found for these configurations in both the hyper-spherical case \cite{bentivegna1} and the flat case \cite{bentivegna2,yoo1,yoo2}. Even taking into account their simplified nature, these lattice models are still more realistic in many aspects than the FLRW models: Firstly, they are locally inhomogeneous in a realistic non-perturbative manner; secondly, the matter content is discrete, rather than a fluid; and thirdly, they are vacuum models, as is the real Universe at almost all points. 

The focus of the present work will be on the surfaces in these models that exhibit reflection symmetry. These include the faces of the cells that constitute the original tiling, as well as a number of the planes that pass through the cell centres. While the geometry of these surfaces is a more complicated problem than those studied previously, it will also enable us to consider more general models, such as the tiling of flat space with an infinite array of cubes, as well as any other initial 2-surface that exhibits reflection symmetry, including models that are not necessarily of the regular lattice type. Our basic requirement will only be that there is an initial 2-surface with reflection symmetry. This immediately implies, in particular, that the symmetry surface itself, as well as its evolution, will be totally geodesic subspaces of the full space-time. 

One feature of particular interest for relativistic cosmological models is the presence (or otherwise) of gravitational radiation. To determine whether a space-time is radiative is not straightforward in general. For our purposes, we exploit the well-known manifestly covariant and non-perturbative electromagnetic analogy for gravity \cite{jantzen}, and say that the flux of gravitational radiation vanishes if the super-Poynting vector vanishes {\edit \cite{Maartens&Bassett:1998, bonilla, gomez}}. We present an argument that this condition implies that the energy carried between cells by the radiation also vanishes, despite the dimensionality of the super-Poynting vector being different from that of an energy flux. Due to the existence of a number of chambers that are entirely enclosed by these reflection symmetric surfaces, we say that our lattice models are ``piecewise silent''.

To implement this program we begin by identifying the geometric quantities that must vanish on reflection symmetric surfaces. We derive the complete (1+2)-dimensional Einstein system along their evolution. Our basic variables for this are the expansion and shear of a set of reference time-like curves, together with the non-vanishing parts of the Weyl tensor. We then evaluate the super-Poynting vector on these surfaces, and consider what this means for the propagation of gravitational waves. Unless otherwise stated, we use Greek letters $\mu$, $\nu$, $\rho$ ... to denote coordinate indices, and Latin letters $a$, $b$, $c$ ... to denote tetrad indices.

\section{Reflection Symmetric Planes in a Lattice Universe}
\label{sec:models}

In this section we will introduce the regular lattice models that we have previously studied in \cite{Clifton_etal:2012,Clifton_etal:2013}. We will then identify the surfaces within them that exhibit reflection symmetry. It will be these surfaces that we consider for the remainder of this paper.

\subsection{A regular lattice of black holes}

The first step in creating a regular lattice model of the Universe is to consider the regular tiling of an abstract 3-dimensional space of constant curvature. If one insists {\edit that} the cells that constitutes this tiling are a set of indentical, regular polyhedra, then there is a finite number of possibilities \cite{coxeter}. These are listed in Table \ref{table1}, below.

If the curvature of the initial reference space is positive, so that it is a hyper-sphere, then there are six possible tilings. These have $5$, $8$, $16$, $24$, $120$ and $600$ cells, which are either tetrahedra, cubes, dodecahedra or octahedra. If the initial space is flat Euclidean 3-space then the only possible regular tiling is with cubic lattice cells. For a negatively curved initial reference space there are four possible tilings, with either cubes, dodecahedra or icosahedra. In both the flat and negatively curved cases the number of cells required to tile the space is inifinite, as we are considering trivial topologies only.

The next step in these constructions is then to place a point-like mass at the centre of each cell\footnote{Alternatively one could place masses on every corner of every cell, but this leads to an identical set of mass distributions.}. To maintain the symmetries of the cells, and in order to have a vacuum space-time, these masses are chosen to be non-rotating and uncharged. One is then in a position to attempt to solve Einstein's equations. No exact solutions for the global geometry of space-time are known for any of the configurations resulting from the tilings in Table \ref{table1}. However, for the six lattices on positively curved spaces it is possible to solve the initial value problem under the assumption of time-symmetry. This has been investigated in some detail in \cite{Clifton_etal:2012}, and even considered for arbitrarily large numbers of irregularly distributed masses in \cite{mik}. The initial value problem in the flat and negatively curved spaces has not been solved exactly, but has been investigated numerically in \cite{bentivegna2,yoo1,bentivegna3}. The exact evolution of some preferred curves has been studied in \cite{Clifton_etal:2013}, and numerical solutions for the evolution of the space-time have been studied in \cite{bentivegna2,yoo2}.

\begin{table}[t!]
\begin{center}
\begin{tabular}{|c|c|c|c|}
\hline
$\begin{array}{c} \bf{Number\;of}\\
  \bf{Cells} \end{array}$  & $\begin{array}{c} \bf{Background}\\
  \bf{Curvature} \end{array}$ & $\begin{array}{c} \bf{Cell}\\
  \bf{Shape} \end{array}$
& $\begin{array}{c} \bf{Lattice}\\
  \bf{Structure} \end{array}$\\ 
\hline
5  & + & Tetrahedron & \{333\} \\
8  & + & Cube & \{433\} \\
16  & + & Tetrahedron & \{334\} \\
24  & + & Octahedron & \{343\} \\
120  & + & Dodecahedron & \{533\} \\
600  & + & Tetrahedron & \{335\} \\
$\infty$  & 0 & Cube & \{434\} \\
$\infty$  & - & Cube & \{435\} \\
$\infty$ & - & Dodecahedron & \{534\}  \\
$\infty$ & - & Dodecahedron & \{535\}  \\
$\infty$  & - & Icosahedron & \{353\} \\
\hline
\end{tabular}
\end{center}
\caption{{\protect{\textit{The regular tilings possible on 3-dimensional spaces of constant positive (+), negative (-) and flat (0) curvature. The lattice structures $\{p,q,r\}$ denote the edges to a cell face, $p$, the number of cell faces that meet at the corner of a cell, $q$, and the number of cells that meet along a cell edge, $r$.}}}}
\label{table1}
\end{table}

\subsection{Reflection symmetric planes}
\label{refplanes}

There are a number of planes in each of the lattices discussed above that exhibit a reflection symmetry. We will classify these planes by considering individual polyhedral cells. Clearly every face of every cell exhibits a reflection symmetry, due to the regularity of these lattices. In addition, however, we can also identify a number of surfaces within each cell that exhibit a reflection symmetry. The precise structure of these internal symmetry surfaces depends on the shape of the cell being considered, but in every case they always pass through the cell centre. We choose to classify these internal surfaces according to whether any cell edges lie within them, or not. In some cases, such as the lattice constructed for icosahedra, the internal symmetry planes all have cell edges lying within them. But in other cases, such as the lattices constructed from cubes, both types of surface exist. We will refer to the symmetry planes containing a cell face as ``f'', the internal planes with cell edges lying in them as ``d'', and the internal planes without edges lying in them as ``p'' \footnote{It should be noted that the surfaces containing cell faces do not necessarily correspond to global symmetry surfaces that contain only cell faces, unless the cell edges are contiguous (as described in \cite{Clifton_etal:2013}). For lattices with non-contiguous edges, such as the 8-cell, the extension of a face surface into any neighbouring cell will coincide with an internal symmetry surface that contains a cell edge. We will still label these reflection surfaces as ``f'', although the reader should keep in mind that globally they may coincide with surfaces labelled ``d''.}.

\begin{figure}[t]
\begin{centering}
\includegraphics[width=7.5cm]{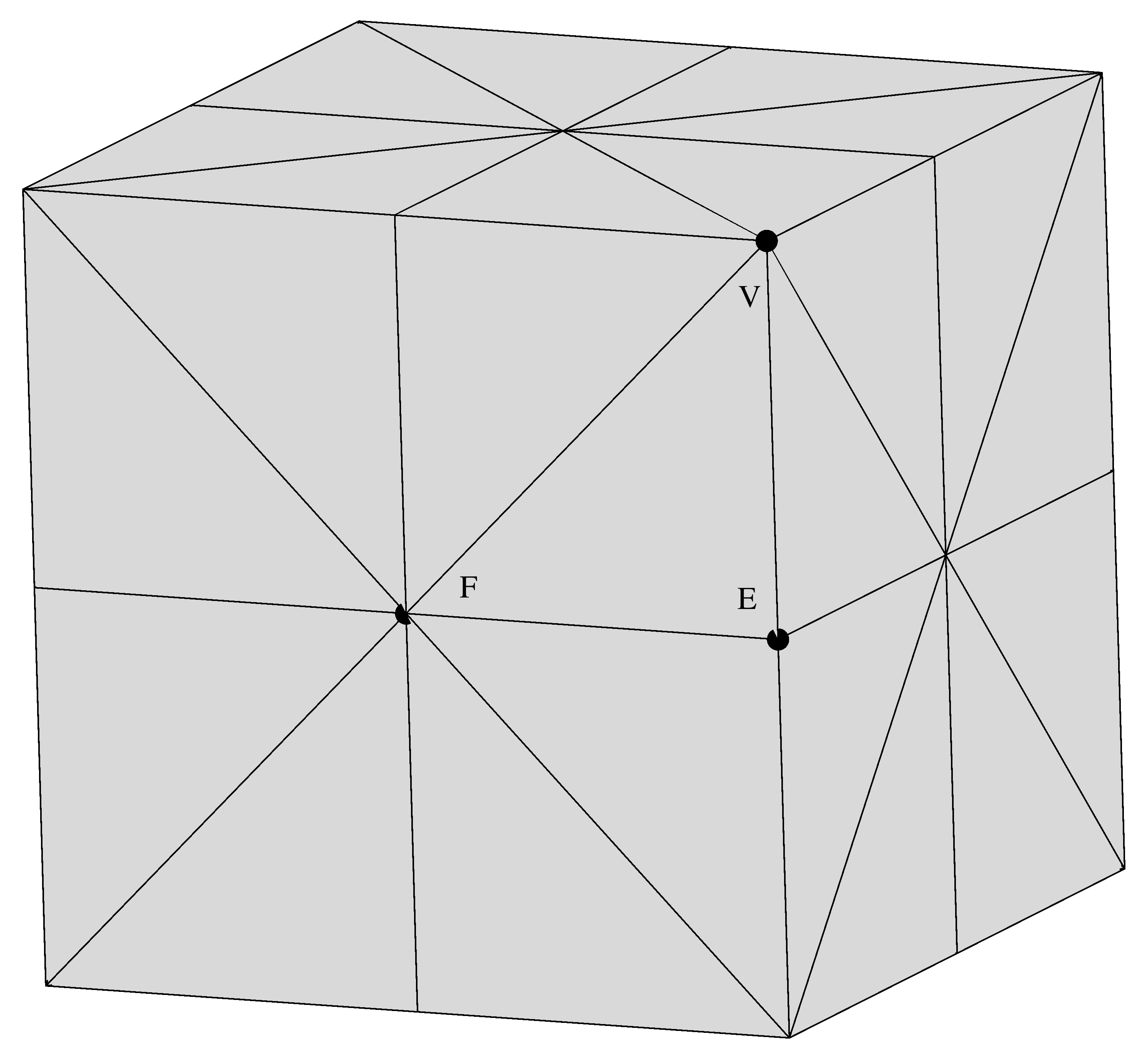} \quad
\includegraphics[width=7.5cm]{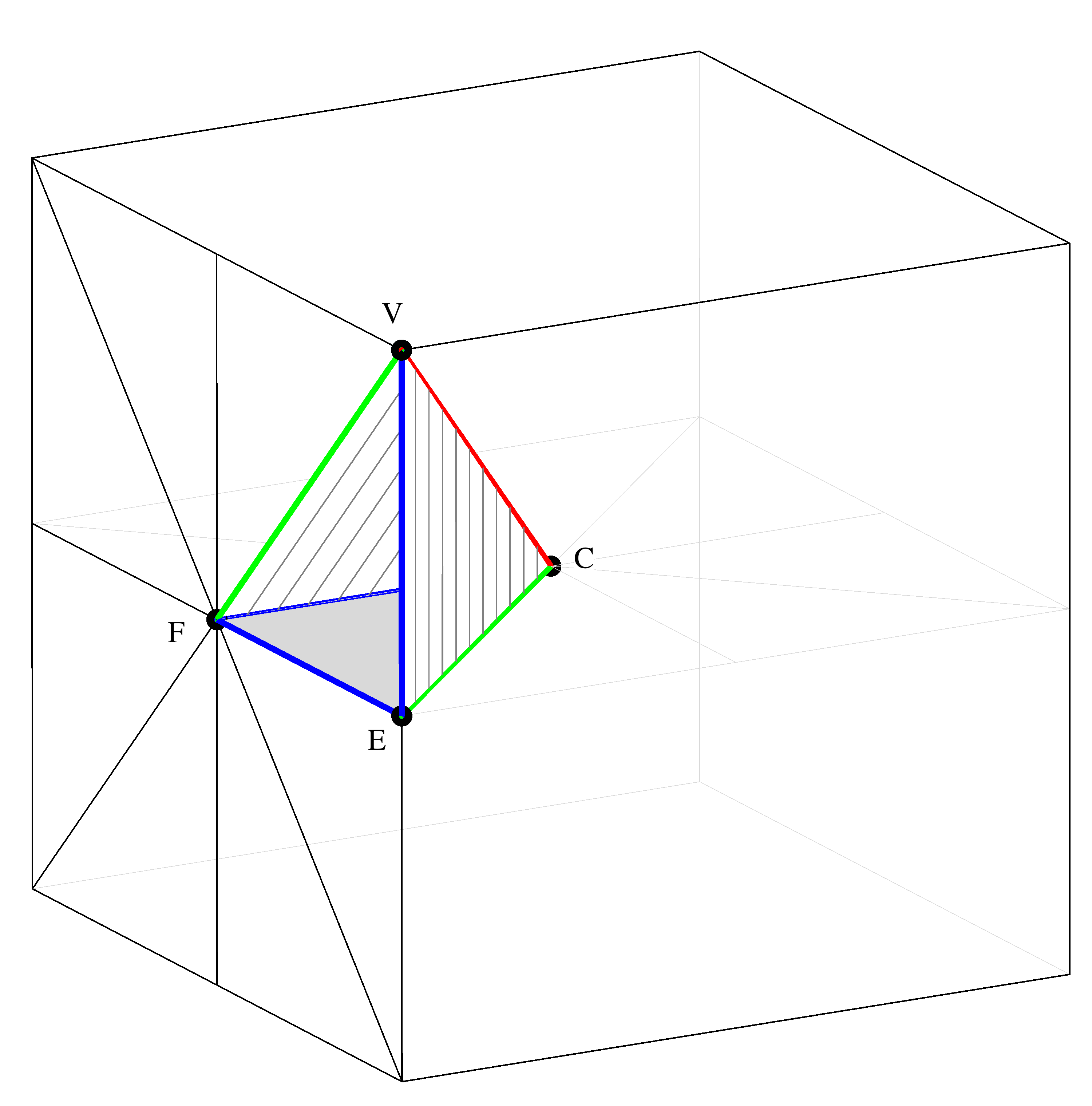}
\par\end{centering}
\caption{An illustration of the tetrahedral chambers that exist within a cubic lattice cell. The diagram on the left shows the triangular faces of the chambers that coincide with the faces of the cell, and the diagram on the right shows one example chamber within the cell. For all chambers, the fourth vertex is the center, C, of the cube. So, for example, the points V, E and F together with C form the vertices of one chamber.}
\centering{}\label{coxeter_complex}
\end{figure}

These reflection symmetric surfaces divide the lattice cells into a number of identical sub-cells, which are called \emph{chambers} \cite{Borovik&Borovik:2010}. We illustrate these chambers, using the example of a cubic lattice cell, in Fig. \ref{coxeter_complex}. In this figure, the chamber vertices are denoted ``V" if they correspond to cell vertex, ``E" if they are at the mid-point of a cell edge, and ``F" if they are at the centre of a cell face. The fourth vertex of each chamber is at the centre of the cell, and is denoted as ``C". Each cubic cell can be seen to consist of 48 chambers, meaning there is a total of 384 chambers in the entire 8-cell lattice. The number of chambers in each lattice cell, and the total number of chambers in each lattice, is given in Table \ref{chambers} for each of the lattice constructed in $S^3$. There are also 48 chambers in each of the cubic lattice cells that exist in $E^3$ and $H^3$, and 120 chambers in each dodecahedral and icosahedral lattice cell in $H^3$. 

Finally, let us consider the symmetries that exist around the curves that constitute the edges of each of the chambers discussed above. These curves are always at the intersection of multiple symmetry surfaces, and so can be classified according to the types of reflection symmetry that they admit. To illustrate this, let us consider the lattice constructed from cubes in $E^3$. In this case there are 6 different distinct chamber edges, which are listed in Table \ref{chamber_edges}, together with their relative length, and the number and type of symmetries that exist around them. If a symmetry is present for an individual cell we refer to its symmetry range as existing for a ``cell'', otherwise, if the whole lattice is required to see the symmetry, we refer to it as existing for a ``lattice''. For the curves in Table \ref{chamber_edges} it can be seen that there exist chamber edges with 2-fold, 3-fold and 4-fold rotational symmetry. This is sufficient to solve for the evolution of each of these curves exactly, using the methods described in \cite{Clifton_etal:2013}.

\section{Implementing a Reflection Symmetry}
\label{sec:imp}

A reflection isometry $\riso$ can be formally defined as follows (see Sec. 9.5 of \cite{rindler}): Let $G$ be a geodesic congruence that is orthogonal to a hypersurface $M$.  For a given point $P$, lying on a geodesic $\gamma \in G$, define $\riso P$ to be the point on the other side of $M$ that lies on the same geodesic $\gamma$, and such that the distance between $P$ and $M$ is the same as the distance between $\riso P$ and $M$. If $\riso$, defined in this way, is an isometry, then we say that $\riso$ is a reflection isometry. It follows that $M$ is fixed under the action of the isometry, such that $\riso M \mapsto M$. We refer to $M$ as a \emph{symmetry surface}. Also, we may assume that $G$ is affinely parametrized in such a way that $\gamma(0) \in M$ for all $\gamma \in G$. A symmetry surface has the useful property of being totally geodesic \cite{rindler}. An example of a reflection symmetry occurs in space-times possessing a moment of time symmetry. In this case the symmetry surface is a spatial hypersurface that has vanishing expansion and shear ({\it i.e.} has zero extrinsic curvature).

As an example of a spatial reflection symmetry we can consider the cell faces in the 8-cell lattice. The locations of the masses at the centre of each lattice cell can then be specified as in Table 3 of \cite{Clifton_etal:2012}. In this case it can be seen that the surface {\edit defined by the polar coordinate} $\phi=\frac{\pi}{4}$ contains four vertices\footnote{{\edit See \cite{Clifton_etal:2012} for an explanation of the $\phi$ coordinate.}}, located at the following positions in Cartesian coordinates in a 4-dimensional Euclidean embedding space:
\begin{eqnarray}
&&\bar V_1=\left( \frac12\,,\, \frac12 \,,\, \frac12\,,\, \frac12\right) \qquad \bar V_2=\left( -\frac12\,,\, \frac12 \,,\, \frac12\,,\, \frac12\right) \nonumber\\
&&\bar V_3=\left( \frac12\,,\, -\frac12 \,,\, \frac12\,,\, \frac12\right) \quad \bar V_4=\left( -\frac12\,,\, -\frac12 \,,\, \frac12\,,\, \frac12\right)\,.
\end{eqnarray}
The spatial metric of the system at the moment of time symmetry is then given by Eq. (14) of \cite{Clifton_etal:2012}, which we can write as
\beq
h_0\,=\,\Phi^4(\chi,\theta,\eta) \left( d\chi^2+\sin^2 \chi \; d\theta^2+\sin^2\chi \; \sin^2\theta \; d\eta^2 \right)\,,
\eeq
where $\eta = \phi - \frac\pi4$, and where $\Phi(\chi,\theta,\eta)$ is given by Eq. (15) of \cite{Clifton_etal:2012}. As $\Phi(\chi,\theta,\eta)$ is an even function of $\eta$, such that $\Phi(\chi,\theta,\eta)=\Phi(\chi,\theta,-\eta)$, the transformation $\eta \to -\eta$ can be seen to be a reflection symmetry. This argument can be straightforwardly generalized to prove that all cell faces in every regular lattice configuration are invariant under reflection, as are the other surfaces identified in Section \ref{refplanes}.

\begin{table}[t]
\begin{center}
\begin{tabular}{|c|c|c|}\hline
  \begin{minipage}{3.8cm} \vspace{4pt} \center
    {\bf Number of Cells in Lattice} \vspace{3pt} \end{minipage} 	
  & \begin{minipage}{4.2cm} \vspace{5pt} \center
    {\bf Number of Chambers per Cell} \vspace{3pt} \end{minipage}
  & \begin{minipage}{3cm} \vspace{4pt} \center {\bf Total Number of Chambers} 
     \vspace{3pt} \end{minipage} \\
\hline
  5   &  24   &   120 \\
   8   &  48   &   384 \\
   16  &  24   &   384 \\
   24  &  48   &  1152 \\
   120 &  120  & 14400 \\
   600 &  24   & 14400 \\
\hline
\end{tabular}
\end{center}
\caption{The number of chambers per cell, and total number of chambers, for each of the lattices constructed in $S^3$.}
\label{chambers}
\end{table}

Now let  $\riso_0$ be a reflection isometry acting on an initial spatial surface $S_0$, with symmetry surface $M_0$. Development by the vacuum Einstein equations guarantees that this isometry is preserved along the evolution \cite{rindler}, and that it extends to a reflection isometry $\riso$ acting on a globally hyperbolic neighborhood $\ghn$ of $S_0$ \cite{Chrusciel:1991, Friedrich&Rendall:2000}.  In particular, for a geodesic slicing, each surface of constant time $S_t$ will be invariant under a reflection symmetry $\riso_t$ with symmetry surface $M_t \subset \ghn$, and there will exist a time-like symmetry surface, $M=\cup M_t$, consisting of fixed points of $\riso$. Moreover, the development of $M_0$ can be shown to define a geodesic congruence on $M$. Intuitively, in a neighbourhood of the symmetry surface, the space-time consists of two identical parts, one on each side of the surface, with identical evolution histories.

\subsection{Constructing a coordinate system, and a tetrad}

We will now explicitly construct a tetrad and a coordinate system that can be used to exploit the reflection symmetry. We start by choosing a coordinate system $(t, x, y, z)$ that is adapted to the reflection symmetry in such a way that $g_{\mu\nu}(t,x,y,z) = g_{\mu\nu}(t,-x,y,z)$. The reflection isometry is then realized as the transformation $x \mapsto -x$, where $x$ can be considered as a (non-unique) reflection parameter, not necessarily along a geodesic. We can then write the metric in terms of functions that are either even or odd with respect to the transformation $x\rightarrow -x$. 

For further specification of the coordinates we take the geodesic slicing $S_t$ to be synchronous, without loss of generality. The symmetry hypersurface $M$ is then orthogonal to $S_t$. We also consider a comoving (time-like) slicing $C_x$ that is adapted to the reflection symmetry so that $C_0=M$ and $\riso C_x = C_{-x}$.  Now let $B_{t,x} = S_t \cap C_x$ denote the spatial \hbox{2-surfaces} formed by the intersections of the $S_t$ and $C_x$ surfaces, and let $(y,z)$ be coordinates that parametrize these surfaces. This completely specifies the coordinate system $(t, x, y, z)$, up to a re-foliation $t \rightarrow t' + f(x,y,z)$, and up to the transformations $x\rightarrow x'(x)$, $y \rightarrow y'(x,y,z)$ and $z \rightarrow z'(x,y,z)$.

\begin{table}[t]
\begin{center}
\begin{tabular}{|c|c|c|c|c|}\hline
  \begin{minipage}{2cm} \vspace{0pt} \center {\bf Chamber Edge} 
     \vspace{0pt} \end{minipage} 
  & \begin{minipage}{2cm} \vspace{0pt} \center {\bf Edge Length}
\vspace{0pt} \end{minipage}
  & \begin{minipage}{2.5cm}  \vspace{5pt} \center {\bf Symmetry Order} 
     \vspace{5pt} \end{minipage}
  & \begin{minipage}{4cm}  \vspace{5pt} \center
    {\bf Reflection Surface Type and Number}  \vspace{5pt} \end{minipage}
  & \begin{minipage}{2.5cm}  \vspace{5pt} \center {\bf Symmetry Range} 
     \vspace{2pt} \end{minipage}  \\
\hline
   EF & 1  & 2  & 1f, 1p  & lattice \\
   VE & 1  & 3  & 3f  & lattice \\
   FC & 1  & 4  & 2p, 2d  & cell \\
   VF & $\sqrt2$  & 2  & 1f, 1d  & lattice \\
   EC & $\sqrt2$  & 2  & 1p, 1d  & cell \\
   VC & $\sqrt3$  & 3  & 3d  & cell  \\
\hline
\end{tabular}
\end{center}
\caption{Properties of the chamber edges for the lattice constructed from cubes in $S^3$. The shortest edge lengths are normalized to unity, and the symmetry order corresponds to the number of reflection symmetric surfaces that intersect an edge. See the text for the notation used for the different types of reflection symmetric surfaces, and an explanation of the symmetry range.}
\label{chamber_edges}
\end{table}

Let us now construct an orthonormal coframe that is adapted to the reflection symmetry. To do this we define $n_\mu = F^{-1} x_{,\mu}$ to be the normalized gradients of the $C_x$ hypersurfaces. The normals $n_\mu$ are then, by construction, perpendicular to the geodesic congruence associated with $S_t$. The two foliations $S_t$ and $C_x$, that are surfaces of constant $t$ and $x$, then give rise to the orthonormal 1-forms $\omega^0:= \td t$ and $\omega^1 := 
F^{-1} \td x$. A coframe can be completed by adding two further normalized 1-forms, which can be written as $\omega^A := P^A \td x + \omega^A{}_K \td x^K$, where we have used the restricted indices $(A,B\ldots= 2,3)$ for frame components and $(K,L\ldots= 2,3)$ for coordinate components. These final two 1-forms have no $t$ component, as we have chosen our coordinate system to be synchronous. The complete orthonormal coframe 
$\omega^a=\omega^{a}{}_{\mu} dx^\mu$ is then given by 
\begin{eqnarray}
\omega^0 &=& \td t \\
\omega^1 &=& \frac1F \td x \\
\omega^A &=& P^A \td x+\omega^{A}{}_{K}\td x^K \, ,
\end{eqnarray}
where we can take $F$ and $\omega^A{}_K$ to be even functions of the coordinate $x=x^1$, while the $P^A$ are odd. 

From the coframe, we can now construct a set of orthonormal frame vectors that are adapted to the symmetry. These are given by
\begin{eqnarray}\label{ONframe1}
   e_0 &=& \partial_t \\ \label{ONframe2}
   e_1 &=& F \partial_x + Q^K \partial_K \\ \label{ONframe3}
   e_A &=& e_A{}^K \partial_K \, ,
\end{eqnarray}
where $(e_A{}^K) = (\omega^A{}_K)^{-T}$, and where $Q^K$ is given by
\begin{equation*}
Q^2=\frac{(-P^2 \omega^{3}{}_{3}+P^3\omega^{2}{}_{3})F}{{\rm det}(\omega^{A}{}_{K})} \qquad {\rm and} \qquad
Q^3=\frac{(P^2 \omega^{3}{}_{2}-P^3\omega^{2}{}_{2})F}{{\rm det}(\omega^{A}{}_{K})}.
\end{equation*}
The $e_A{}^K$ in Eq. (\ref{ONframe3}) are even functions of $x$, while the functions $Q^K$ in Eq. (\ref{ONframe2}) are odd. This completes our specification of the tetrad, which is now uniquely defined up to a rotation of the vectors $e_2$ and $e_3$.

\subsection{Tetrad commutation functions}

The tetrad vectors in Eqs. (\ref{ONframe1})-(\ref{ONframe3}) can be used to define a set of commutation function, $\gamma^{a}_{\phantom{a} bc}$, via
\begin{equation}
 [e_a,\,e_b] = \gamma^c{}_{ab} \,e_c \, .
\end{equation}
It follows immediately from Eq. (\ref{ONframe3}) that
\begin{equation}\label{gamma1}
  \gamma^{1}_{\phantom{1}AB} =0 \ .
\end{equation}
Furthermore, as the vector $e_0$ is geodesic and hyper-surface orthogonal we immediately know that it must be irrotational, so that
\begin{equation}\label{gamma1b}
  \gamma^{0}_{\phantom{0}A 1} =0 =\gamma^{0}_{\phantom{0}0 1}\ .
\end{equation}
These relations are a consequence of the choice of frame, and apply to any frame that is adapted to hypersurfaces in the way we have specified. They do not depend on any symmetry properties about those hypersurfaces.

A commutator on which the reflection symmetry does have an effect is
\begin{equation*}
[e_1\,,\,e_A] = F e_A{}^K{}_{\!\!,x} \,\partial_K+Q^L e_A{}^K{}_{\!\!,L}\,\partial_K-e_A{}^L F_{,L} \,\partial_x-e_A{}^L \, Q^K{}_{,L}\,\partial_K\,.
\end{equation*}
Evaluating this expression on the symmetry surface we see that only the third term can be non-zero. This implies that on the symmetry surface we have
\begin{equation}
\label{gamma2}
  \gamma^{A}_{\phantom{A} 1B} = 0\,.
\end{equation}
Similarly, if we consider the commutators
\begin{equation*}
[e_1\,,\,e_0] =  -F_{,t} \partial_x - Q^K_{\phantom{K} ,t} \partial_K     \,  \qquad {\rm and} \qquad [e_0\,,\,e_A] = e_{A \phantom{K} ,t}^{\phantom{A} K} \partial_K \, ,
\end{equation*}
then we see that the second term in the right hand side of the first equation vanishes. This implies that on the symmetry surface we also have
\begin{equation}
\label{gamma2b}
  \gamma^{A}_{\phantom{A} 10} = 0 =\gamma^{1}_{\phantom{1} 0A} \,.
\end{equation}

Eqns. (\ref{gamma1})-(\ref{gamma2b}) can be summarized as follows: {\it Commutation functions with an odd number of indices equal to 1 are identically zero on the reflection surface}. The same result then follows for the Ricci rotation coefficients, which can be defined by
\begin{equation}\label{RRC}
\Gamma_{abc} = \frac{1}{2} \left( \gamma_{acb} + \gamma_{bac} - \gamma_{cba} \right).
\end{equation}
If, following Ellis \& MacCallum \cite{SKB}, we define $n^{\alpha \beta} := \frac{1}{2} \gamma^{(\alpha}_{\phantom{(\alpha} \gamma \delta}  \epsilon^{\beta) \gamma \delta}$ and $a_{\beta} := \frac{1}{2} \gamma^{\alpha}_{\phantom{\alpha} \beta \alpha}$, then it can be seen that Eqs. (\ref{gamma1}) and (\ref{gamma2}) imply
\begin{equation}
n_{11}= n_{22} = n_{33} =n_{23} =a_1  =0.
\end{equation}
The only independent non-zero parts of $n_{\alpha \beta}$ and $a_{\alpha}$ on the symmetry surface are therefore $n_{12}$, $n_{13}$, $a_2$ and $a_3$. This is a significant simplification.

\subsection{Kinematic quantities}

Let us use the notation $u^{\mu}=e_0^{\phantom{0} \mu}$. We can then define a projection tensor $h_{\mu \nu} = g_{\mu \nu} + u_{\mu} u_{\nu}$, and perform an irreducible decomposition of $u^{\mu}$ such that
\begin{equation}
u_{\mu ; \nu} = - u_{\nu} \dot{u}_{\mu} + \sigma_{\mu \nu} + \frac{1}{3} \Theta h_{\mu \nu} +\omega_{\mu \nu} \, ,
\end{equation}
where the over-dot denotes differentiation along $u^{\mu}$, such that $\dot{X} = u^{\mu} X_{;\mu}$. The tensors $\sigma_{\mu \nu}$ and $\omega_{\mu \nu}$ are the symmetric and anti-symmetric parts of the projected trace-free part of $u_{\mu ; \nu}$, respectively. The remaining variables are the expansion scalar $\Theta = u^{\mu}_{\phantom{\mu} ; \mu}$, and the acceleration vector $\dot{u}^{\mu}$. Collectively, these variables are known as the kinematic quantities associated with $u^{\mu}$.

We have already chosen $S_t$ to be a geodesic slicing, which means that $\dot{u}^{\mu}=0$. We have also defined $u^{\mu}$ as being orthogonal to a set of spatial hypersurfaces, which automatically means $\omega_{\mu \nu}=0$. Again, these results are true by construction, and not because of the reflection symmetry.
We can see from Eq. (\ref{gamma2b}), however, that the reflection symmetry does imply that on the symmetric surface we have in tetrad components that
\begin{equation}
\sigma_{12}=\sigma_{13} = \Omega_2 = \Omega_3 =0\, ,
\end{equation}
where $\Omega^{\alpha} = \frac{1}{2} \eta^{\alpha \beta \gamma} \dot{e}_{\gamma} \cdot e_{\beta}$ is the angular velocity of the triad vectors $e^{\alpha}$ in the rest-frame of an observer with 4-velocity $u^{\mu}$. The only independent non-zero kinematic quantities on the reflection symmetric surface are therefore $\sigma_{22}$, $\sigma_{23}$, $\sigma_{33}$ and $\Theta$. The remaining angular velocity component, $\Omega_1$, can be chosen freely, and may be used to set either $\gamma^{3}_{\phantom{3} 02}$ or $\gamma^{2}_{\phantom{2} 03}$ to zero (but not both).

\subsection{Weyl tensor}
\label{sec:weyl}

The remaining quantities that need to be considered in our system are the components of the Weyl tensor, which in vacuum are given by
\cite{uggla2}
\beq \label{weyl}
\hspace{-10pt}
C^{a}{}_{bcd}\,=\,R^{a}{}_{bcd}\,=\,e_c(\Gamma^{a}{}_{bd})-e_d(\Gamma^{a}{}_{bc})+\Gamma^{a}{}_{ec}\Gamma^{e}{}_{bd}-\Gamma^{a}{}_{ed}\Gamma^{e}{}_{bc}-\Gamma^{a}{}_{be}\gamma^{e}{}_{cd} \, ,
\eeq
where the $\Gamma^a{}_{bc}$ are the Ricci rotation coefficients defined in Eq. (\ref{RRC}). This tensor can be decomposed with respect to $u^a$ into electric and magnetic parts, as follows \cite{cosmological}:
\begin{eqnarray} \label{E}
&&E_{ac}\,:=\,C_{abcd}u^b u^d \, ,\\
&&H_{ac}\,:=\, \frac12 \eta_{ab}{}^{ef} C_{efcd}u^b u^d\, .  \label{H}
\end{eqnarray}
Inserting into these definitions the simplifications obtained from Eqs. (\ref{gamma1}) and (\ref{gamma2}) then implies that on the symmetry surface we have
\begin{eqnarray}
&&E_{12}\,\equiv\, 0\,, \qquad E_{13}\,\equiv\, 0 \,, \qquad  H_{11}\,\equiv\, 0 \,, \\
&& H_{22}\,\equiv\, 0 \,, \qquad H_{23}\,\equiv\, 0\,, \qquad H_{33}\,\equiv\, 0\, .
\end{eqnarray}
The only non-zero parts of the Weyl tensor on the symmetry boundary are therefore $E_{11}$, $E_{22}$, $E_{33}$, $E_{23}$, $H_{12}$ and $H_{13}$. This immediately implies that
\beq
E_{ab}H^{ab} \equiv 0\, .
\eeq
The scalar $E_{ab}H^{ab}$  is an observer independent quantity, unlike $E_{ab}$ and $H_{ab}$ themselves, and therefore has a special physical significance. That it vanishes is the first physically important implication of the reflection symmetry that we have been investigating.

If we now define the projection tensor
\beq
N^{a}{}_{b}:=h^{a}{}_{b}-n^an_b=\delta^{a}{}_{b}+u^a u_b -n^a n_b\,,
\eeq
where $n^a = e_1^{\phantom{1} a}$ is the normal to the reflection surface, then the scalar, vector and projected symmetric trace-free spatial 2-tensorial components of a spatial symmetric trace-free tensor $\psi_{ab}$ can be written as \cite{clarkson}
\begin{eqnarray}
\fscalar{\psi} &\,:=\,& n^a n^b \psi_{ab} \\ 
\fvec{\psi}_a &\,:=\,& N_{a}{}^{b} n^c \psi_{bc} \\
\ftensor{\psi}_{ab} &\equiv& \psi_{ \{ab  \}} \,:=\, \bigl(N_{(a}{}^{c} N_{b)}{}^{d}-\frac12 N_{ab} N^{cd}\bigr)\psi_{cd}\,.
\end{eqnarray}
In this notation the results above can be expressed in the following way:
\begin{eqnarray}
&&\fvec{E}_2\,\equiv\, 0\,, \qquad \fvec{E}_3\,\equiv\, 0  \\
\label{magneto}
&&\fscalar{H}\,\equiv\, 0 \,, \qquad \ftensor{H}_{22}\,\equiv\, 0 \,, \qquad \ftensor{H}_{33}\,\equiv\, 0\,.
\end{eqnarray}
For reference, we note the well-known decomposition of $\psi_{ab}$ in [33]
\begin{eqnarray}\label{param}
\label{split}
   &&\psi_+ := -\frac32 \psi_{11} = \frac32(\psi_{22}+\psi_{33}) \ ,\quad
   \psi_-  := \frac{\sqrt3}2 (\psi_{22}-\psi_{33}) \\
   &&\psi_1 := \sqrt3 \psi_{23} \ , \quad \psi_2 := \sqrt3 \psi_{31} \ ,\quad
   \psi_3  := \sqrt3 \psi_{12}\,,
\end{eqnarray}
leading to the relations
\begin{eqnarray}
&& \fscalar{\psi} \,=\,-\frac{2}{3} \psi_+  \, ,\\
&& \fvec{\psi}_2 \,=\, \frac {1}{\sqrt3} \psi_3 \,, \qquad \fvec{\psi}_3 \,=\, \frac {1}{\sqrt3} \psi_2  \,, \\
&&  \psi_{\{ 22  \}}\,=\, \frac {1}{\sqrt3}  \psi_{-}  \,, \qquad  \psi_{\{ 23  \}}\,=\, \frac {1}{\sqrt3}  \psi_{1} \, .
\end{eqnarray}
The only non-zero components of $H_{ab}$ are then given by
\beq 
\fvec{H}_2 \,=\, \frac {1}{\sqrt3} H_3 \,, \qquad \fvec{H}_3 \,=\, \frac {1}{\sqrt3} H_2\,,
\eeq 
which, defined in the way they have been, lie in the reflection symmetric surface. The electric part of the Weyl tensor, on the other hand, has no vector components in the reflection symmetric surface.

\section{Evolution of Reflection Symmetric Surfaces}

Let us now consider the geometry of the $2+1$-dimensional symmetry hypersurface $M$. The metric tensor of this space is
\begin{equation}
\gamma_{\mu \nu} := g_{\mu \nu} - n_{\mu} n_{\nu},
\end{equation}
where $n^{\mu}$ is the space-like unit vector orthogonal to $M$ that we considered above. Now, the time-like vector field $u^{\mu}$ is, by construction, orthogonal to $n^{\mu}$, and so is already projected ({\it i.e.} $u^{\mu} = \gamma^{\mu}_{\phantom{\mu} \nu} u^{\nu})$. The projected covariant derivative of $u^{\mu}$ is
\begin{eqnarray} 
D_{\mu} u^{\nu} &:=& \gamma^{\rho}_{\phantom{\rho} \mu} \gamma^{\nu}_{\phantom{\nu} \sigma} \nabla_{\rho} u^{\sigma} =  \gamma^{\rho}_{\phantom{\rho} \mu} \nabla_{\rho} u^{\nu} - n^{\nu} u^{\rho} K_{\mu \rho} \, ,
\end{eqnarray}
where $K_{\mu \nu} = - \gamma^{\rho}_{\phantom{\rho} \mu} \gamma^{\sigma}_{\phantom{\sigma} \nu} \nabla_{\rho} n_{\sigma}$ is the extrinsic curvature of $M$. This expression can be irreducibly decomposed as
\begin{equation} \label{ID2}
D_{\mu} u_{\nu} = -u_{\mu} \dot{u}_{\nu} + \varsigma_{\mu \nu} + \frac{1}{2} \theta \gamma_{\mu \nu} + \varpi_{\mu \nu} \, ,
\end{equation}
where $\varsigma_{\mu \nu} = \varsigma_{(\mu \nu)}$ and $\varpi_{\mu \nu} = \varpi_{[\mu \nu]}$ are the shear and vorticity tensors in $M$, defined such that $\varsigma_{\mu \nu} u^{\nu} = 0 =\varpi_{\mu \nu} u^{\nu}$ and $\varsigma^{\mu}_{\phantom{\mu}\mu}=0$. They measure the volume preserving deformation and the rotation of $u^{\mu}$ in the hypersurface $M$, respectively. The expansion of $u^{\mu}$ in this space is given by the scalar $\theta = D_{\mu} u^{\mu}$.

Similarly, the projected second covariant derivative of $u^{\mu}$ is given by
\begin{eqnarray}
D_{\mu} D_{\nu} u^{\rho} &:=& \gamma^{\sigma}_{\phantom{\sigma} \mu} \gamma^{\tau}_{\phantom{\tau} \nu} \gamma^{\rho}_{\phantom{\rho} \phi} \nabla_\sigma \left( \gamma^{\chi}_{\phantom{\chi} \tau} \gamma^{\phi}_{\phantom{\phi} \psi} \nabla_{\chi} u^{\psi} \right) \\
&\, =& \gamma^{\sigma}_{\phantom{\sigma} \mu} \gamma^{\tau}_{\phantom{\tau} \nu} \gamma^{\rho}_{\phantom{\rho} \phi} \nabla_{\sigma} \nabla_{\tau} u^{\phi} + K_{\mu}^{\phantom{\mu} \rho} u^{\phi} K_{\nu \phi} + \gamma^{\rho}_{\phantom{\rho} \tau} K_{\mu \nu} n^{\phi} \nabla_{\phi} u^{\tau} \, .
\end{eqnarray}
This can be used to write down the following expression for Riemann tensor of the $1+2$-dimensional space:
\begin{eqnarray}
\mathcal{R}^{\mu \nu}_{\phantom{\mu \nu} \rho \sigma}  u_{\mu} &:=& 2 D_{[\sigma} D_{\rho]} u^{\nu}\\
&\, =& 2 \gamma^{\tau}_{\phantom{\tau} \sigma} \gamma^{\phi}_{\phantom{\phi} \rho} \gamma^{\nu}_{\phantom{\nu}\chi} \nabla_{[\tau} \nabla_{\phi]} u^{\chi} + 2 K^{\nu}_{\phantom{\nu} [\sigma} K_{\rho ] \tau} u^{\tau}.
\end{eqnarray}
Seeing that the first term in this expression contains the definition of the Riemann tensor of the $4$-dimensional space-time, and recognising that this expression must be true for any vector that lies in $M$, gives the Gauss embedding equation for $M$:
\begin{equation} 
\mathcal{R}_{\mu \nu \rho \sigma} +K_{ \mu \sigma} K_{\nu \rho} - K_{\mu \rho} K_{\nu \sigma} =  \gamma^{\tau}_{\phantom{\tau} \mu}  \gamma^{\phi}_{\phantom{\phi} \nu}  \gamma^{\chi}_{\phantom{\chi} \rho}  \gamma^{\psi}_{\phantom{\psi} \sigma} R_{\tau \phi \chi \psi} \, .
\end{equation}
As the hypersurface $M$ is reflection symmetric, we must have symmetry under $n^{\mu} \rightarrow -n^{\mu}$. This means that $K_{\mu \nu}=0$, and the Riemann tensor of $M$ is simply given as the projection of the Riemann tensor of the space-time:
\begin{equation} \label{gauss}
\mathcal{R}_{\mu \nu \rho \sigma}  =  \gamma^{\tau}_{\phantom{\tau} \mu}  \gamma^{\phi}_{\phantom{\phi} \nu}  \gamma^{\chi}_{\phantom{\chi} \rho}  \gamma^{\psi}_{\phantom{\psi} \sigma} R_{\tau \phi \chi \psi} \, .
\end{equation}
We can use this expression to calculate the constraint and evolution equations for the kinematic quantities defined in Eq. (\ref{ID2}).

\subsection{Effective fluid description}

Contracting Eq. (\ref{gauss}), and using the fact that the full space-time is a vacuum solution of Einstein's equations, so that $R_{a b}=0$, gives the following expressions for the tetrad components of the Ricci tensor of $M$:
\begin{eqnarray*}
\mathcal{R}_{00} &= -E_{11} \, , \qquad \qquad 
\mathcal{R}_{02} &= H_{13} \, ,\\
\mathcal{R}_{03} &= -H_{12}\, , \qquad \qquad
\mathcal{R}_{22} &= E_{33}\, ,\\
\mathcal{R}_{23} &= -E_{23}\, , \qquad \qquad
\mathcal{R}_{33} &= E_{22}\, ,
\end{eqnarray*}
where we have used Eqs. (\ref{weyl}), (\ref{E}) and (\ref{H}) to write $R_{a b c d}$ in terms of $E_{a b}$ and $H_{a b}$. It can immediately be seen that $\mathcal{R} := \gamma^{a b} \mathcal{R}_{a b}=0$.

These equations can be seen to satisfy a lower-dimensional version of Einstein's equations, $\mathcal{R}_{a b} - \frac{1}{2} \gamma_{ab} \mathcal{R} = \mathcal{T}_{ab} $, with an effective energy-momentum given by
\begin{equation}
\mathcal{T}_{a b} = (\tilde{\rho} +\tilde{p}) u_a u_b + \tilde{p} \gamma_{ab}  + \tilde{\pi}_{ab}  + \tilde{q}_{a} u_b + u_a \tilde{q} _b ,
\end{equation}
where 
\begin{eqnarray*}
\tilde{p} &=& \frac{1}{2} \tilde{\rho} = -\frac{1}{2} E_{11} \, ,  \quad   \quad \tilde{q}_{a} = \left( 0,  -H_{13} , H_{12} \right) \, ,\\
\tilde{\pi}_{a b} &=& 
\left( \begin{array}{ccc}
0 & 0 &0 \\
0 & -\frac{1}{2} (E_{22}-E_{33} ) & -E_{23}  \\
0 & -E_{23} & \frac{1}{2} (E_{22}-E_{33} )  \end{array} \right) \, .
\end{eqnarray*}
This effective fluid can be seen to have an effective equation of state $w := {\tilde{p}}/{\tilde{\rho}} =1/2$, and an effective heat flow, $\tilde{q}_{a}$, that is non-zero if and only if $H_{ab}$ is non-zero on $M$. The effective anisotropic pressure, $\tilde{\pi}_{ab}$, is non-zero at any points where $E_{ab}$ is not symmetric under a local spatial rotation in $M$.

We emphasize that in no way does the effective fluid we have defined above constitute an actual matter field in the space-time. The space-time is vacuum, but in the dimensionally reduced system the intrinsic geometry behaves as though it satisfies Einstein's equations with a matter source. This effective matter source is completely determined by the Weyl tensor of the $4$-dimensional space-time.

\subsection{Expansion of the reflection symmetric surfaces}

We can now perform a $1$+$2$-dimensional covariant decomposition of the lower dimensional gravitational system, in order to find the equations that govern the expansion of our reflection symmetric surfaces. Covariant studies of lower-dimensional systems such as this have been performed in the past \cite{doug}, but as far as we are aware they have not been performed with a general fluid content (including heat flow and anisotropic pressure). The full set of covariant equations are therefore presented in the appendix.

In order to discuss the evolution of our reflection symmetric surfaces, we define their areal scale factor $a$ via
\begin{equation}
\theta = 2 \frac{\dot{a}}{a} \, .
\end{equation}
The first and second time derivatives of this scale factor can then be deduced from the equations in the appendix, and are given in our case by
\begin{eqnarray}
\label{Friedmann1}
\frac{\dot{a}^2}{a^2} &=& \tilde{\rho} + \varsigma^2 - \mathcal{K} \, , \\
\label{Friedmann2}
\frac{\ddot{a}}{a} &=&  - \frac{\tilde{\rho}}{2} -\varsigma^2 \, ,
\end{eqnarray}
where $\varsigma^2 = \frac{1}{2} \varsigma_{\mu \nu} \varsigma^{\mu \nu}$, and $\mathcal{K}$ is the Gaussian curvature\footnote{Defined such that the Ricci scalar of these surfaces is ${}^{(2)}\mathcal{R}=2 \mathcal{K}$.} of the $2$-dimensional surfaces orthogonal to $u^{\mu}$. The evolution equations for $\tilde{\rho}$, $\varsigma^2$ and $\mathcal{K}$ are then given by
\begin{eqnarray}
\label{Friedmann3}
\dot{\tilde{\rho}} + 3 \frac{\dot{a}}{a} \tilde{\rho} &=& -Q_1 - Q_2 \, ,\\
\label{Friedmann4}
\hspace{-10pt}(\varsigma^2)\dot{{}} +4 \frac{\dot{a}}{a} \varsigma^2 &=& Q_1  \, , \\
\label{Friedmann5}
\hspace{-35pt} (-\mathcal{K})\dot{{}} + 2 \frac{\dot{a}}{a}  (-\mathcal{K}) &=& Q_2 \, ,
\end{eqnarray}
and where we have defined $Q_1 :=  \varsigma^{\mu \nu} \tilde\pi_{\mu \nu}$ and $Q_2 := D^{\mu} \tilde q_{\mu}$. 

The set of Eqs. (\ref{Friedmann1})-(\ref{Friedmann5}) are very similar to the $3$+$1$-dimensional Friedmann equations, with $\tilde{\rho}$ behaving like pressureless  dust, $\varsigma^2$ behaving like radiation, and $\mathcal{K}$ behaving like a spatial curvature term. In this analogy, the $Q_1$ and $Q_2$ terms in Eqs. (\ref{Friedmann3})-(\ref{Friedmann5}) take the place of energy exchange terms between the dust and radiation, and between the dust and spatial curvature, respectively. Such systems have already been studied in the context of FLRW cosmologies (see e.g. \cite{ex}). It is interesting to note that $\tilde{\rho}$ behaves like a dust term, even though its effective equation of state in the lower-dimensional system is $w=1/2$.

If the $Q_1$ and $Q_2$ terms are non-zero then we will require evolution equations for them in order to close the system. The evolution equation for $Q_2$ can be found from the equations in the appendix, and is in general a PDE. This means that in general the expansion of the reflection symmetric surfaces cannot be solved independently at each point in space. This is not surprising as $Q_2$ is determined by the magnetic part of the $4$-dimensional Weyl tensor. To find an evolution equation for $Q_1$ we require information from the full $3$+$1$-dimensional system, as in the dimensionally reduced system this equation corresponds to the evolution equation for the effective anisotropic pressure. Such an equation cannot be found without knowledge of an effective equation of state, which is absent from the $2$+$1$-dimensional system.

\section{The Propagation of Gravitational Waves}

Let us now consider the propagation of gravitational waves in space-times with reflection symmetric surfaces. To do this it is instructive to consider the gravitational analogue of the Poynting vector\footnote{In electromagnetism the Poynting vector corresponds to the flux density of energy in the electromagnetic field.}, which can be defined as \cite{Maartens&Bassett:1998}
\beq
P_a\,=\, \epsilon_{abc}E^{b}{}_{d}H^{cd} \ .
\eeq
This quantity is, in fact, best referred to as the ``super-Poynting vector'' of the gravitational field, as it is derived from the Bel-Robinson tensor, which acts like the super-energy-momentum tensor of the free gravitational field \cite{Senovilla:2000}. As such, the $4$-vector  $P_a$ does not have the dimensionality of an energy flux density, and therefore requires some interpretation if it is to be used as a criterion {\edit for the non-existence of a flux of gravitational waves} \cite{Zakharov:1973}. We will do this in Section \ref{sec:spwf}, {\edit and will use the results to motivate our upcoming definition of ``piecewise silence'' in Section \ref{sec:pws}.}

\subsection{The super-Poynting vector for weak fields}
\label{sec:spwf}
In the standard linearized theory of weak-field gravitational waves the starting point is to write the metric as a perturbation of the Minkowski space-time
\begin{equation}
   g_{\mu\nu} = \eta_{\mu\nu} + \h_{\mu\nu} \, ,
\end{equation}
where $\h_{\mu\nu}$ is small in the sense that
\begin{equation}
   |\h_{\mu\nu}| \ll 1 \, ,
\end{equation}
and that additional smallness requirements on the derivatives of $\h_{\mu\nu}$ are satisfied (see e.g.\ \cite{Schutz:2009}). The Weyl tensor is then given by
\begin{equation}
  C_{\mu\nu\rho\sigma} = \h_{\mu[\sigma,\rho]\nu}
                        -\h_{\nu[\sigma,\rho]\mu}\ ,
\end{equation} 
and we can choose a gauge such that $\eta^{\mu\nu} \h_{\mu\nu}=0$ and $\h_{\mu\nu} =A_{\mu\nu} \sin \phi$, where $\phi=k_\mu x^\mu$. If we now choose the $4$-velocity of the observer to be $\bar u^\mu= (1,0,0,0)$, and choose the radiation to propagate along the $z$-axis along $\bar n^\mu= (0,0,0,1)$, then the tangent $4$-vector to the ray is $k^\mu = (\omega, 0, 0, \omega) = \omega( \bar u^\mu + \bar n^\mu)$, where $\omega$ is the angular frequency. If we now define a screen space metric by $s_{\mu\nu} = \eta_{\mu\nu}+\bar u_\mu \bar u_\nu - \bar n_\mu \bar n_\nu$, then we can use a $1$+$1$+$2$-decomposition with respect to $\bar u^\mu$ and $\bar n^\mu$ to write\footnote{Note that $A_+:= A_{11}$ follows the standard convention in the gravitational wave literature, and differs by a factor $-\frac32$ from the parametrization introduced in Eq. (\ref{param}).}
\begin{equation}
   A_{\mu\nu} = A_{\{\mu\nu\}} = \left( \begin{array}{cccc}
                                 0  & 0         & 0         & 0 \\
                                 0  & A_+       & A_\times  & 0 \\
                                 0  & A_\times  & -A_+      & 0 \\
                                 0  & 0         & 0         & 0 \end{array}
                               \right)= A_+ I_+ + A_\times I_\times            \, ,                        
\end{equation}
where $I_+$ and $I_\times$ are unit linear-polarization matrices defined by \cite{Misner_etal:1973}
\begin{equation}
   I_+ := \left( \begin{array}{cccc}
           0  & 0  & 0   & 0 \\
           0  & 1  & 0   & 0 \\
           0  & 0  & -1  & 0 \\
           0  & 0  & 0   & 0 \end{array} \right) \ ,\qquad
   I_\times := \left( \begin{array}{cccc}
           0  & 0  & 0  & 0 \\
           0  & 0  & 1  & 0 \\
           0  & 1  & 0  & 0 \\
           0  & 0  & 0  & 0 \end{array}  \right) \, ,
\end{equation}
such that $(I_+)^2=I_{\rm S}$ and $(I_\times)^2=I_{\rm S}$, where $I_{\rm S}$ is the screen space unit matrix. The electric and magnetic parts of the Weyl tensor then take the form
\begin{eqnarray}
   E_{\mu\nu} &=& \frac{1}{2}\omega^2 \sin\phi A_{\mu\nu} \, ,\\
   H_{\mu\nu} &=& \rot_\mu{}^\rho E_{\rho\sigma} \rot_\nu{}^\sigma
              = \epsilon_\mu{}^\rho E_{\rho\nu} \, ,
\end{eqnarray}
where
\begin{equation}
   \rot := \left( \begin{array}{cccc}
          1  &  0               & 0               &  0 \\
          0  &  \frac1{\sqrt2}  & \frac1{\sqrt2}  &  0 \\[3pt]
          0  & -\frac1{\sqrt2}  & \frac1{\sqrt2}  &  0 \\
          0  &  0               & 0               &  1    \end{array} \right)
\end{equation}
is the rotation matrix corresponding to a rotation by $\pi/4$ in the screen space, and
\begin{equation}
   \epsilon_{\mu\nu} = \bar n^\rho \bar u^\sigma \epsilon_{\mu\nu\rho\sigma}
\end{equation}
is the 2-dimensional Levi-Civita tensor in the screen space. The electric and magnetic Weyl tensor amplitudes can then be expressed in matrix form as
\begin{eqnarray}
   A_E &=& A_+ I_+ +  A_\times I_\times \, , \\ 
   A_H &=& \rot \half A_E \rot^{\rm T} = A_\times I_+ - A_+ I_\times \, ,
\end{eqnarray}
so that the super-Poynting vector becomes $P_\mu = (0,0,0,P_z)$ where
\begin{equation}\label{superflux}
   P_z = \omega^4 \sin^2 \phi (A_+^2 + A_\times^2) \, .
\end{equation}
This can be compared with the energy flux density \cite{Misner_etal:1973}
\begin{equation}
    q_\mu = \frac1{32\pi} \,\omega^2 (A_+^2 + A_\times^2) \, \delta^z_\mu \, ,
\end{equation}
and leads to the following expression for the energy flux in terms of the super-Poynting vector:
\begin{equation}\label{fluxequality}
    q_\mu = \frac1{16\pi} \omega^{-2} \bar P_\mu  \, ,
\end{equation}
where $\bar P_\mu$ denotes the average of $P_{\mu}$ over one period. Although the relationship between the super-Poynting vector and gravitational radiation has been considered many times in the literature \cite{Maartens&Bassett:1998,bonilla,gomez}, this is to the best of our knowledge the first time that it has been directly related to the energy flux of weak-field gravitational waves. From the physical point of view, Eq. (\ref{fluxequality}) indicates that the super-Poynting flux can be interpreted as being proportional to the energy flux density, with a proportionality factor that depends on the frequency of the wave. This makes it seem reasonable to use $P_\mu$ as an indicator of the direction of energy flux in gravitational waves, even if understanding the magnitude of that flux requires a more elaborate treatment. {\edit In what follows we will do just this, and use the vanishing of $P_{\mu}$ as an indicator for the vanishing of the flux of gravitational radiation.}

\subsection{Piecewise silent universes}
\label{sec:pws}
A special class of cosmological solutions of Einstein's equations are the \lq\lq Silent universes\rq\rq. These solutions are obtained under the assumptions of (i) irrotational dust and (ii) a vanishing magnetic part of the Weyl tensor. These solutions are not assumed to admit any symmetries {\it ab initio}, but are simple enough to be able to be studied using a dynamical system approach.  The non-vanishing variables for these space-times are the the expansion scalar, the shear tensor, the electric part of the Weyl tensor, the energy density in dust, and the cosmological constant:
\beq
\lbrace \theta, \sigma_+,\sigma_-, E_+, E_-, \Lambda, \rho \rbrace\,.
\eeq
As $H_{ab}$ vanishes by assumption, it is usually said that there are no gravitational waves in these space-times. In terms of the discussion above, this can be stated as a sufficient condition for the super-Poynting vector to vanish everywhere, such that there is no flux of gravitational waves between any two points in space-time.

If we want to deepen our understanding of the silent properties of the real Universe it seems necessary that we should relax the assumption of $H_{ab}=0$ globally, and that we could instead introduce a local notion of silence. In this spirit, and following the discussion above, we define a universe to be  piecewise silent if
\begin{enumerate}
\item
There is a well-defined subdivision of the universe into two or more non-overlapping (except for the boundaries) spatial regions.
\item
All observers comoving with the boundaries to these regions measure $P_a$ to have no component perpendicular to the boundary.
\end{enumerate}
This definition can be seen to be satisfied for the subdivision into chambers of our lattice universe, if we use the results of Section \ref{sec:weyl} and the subdivision described in Sec. \ref{sec:models}. Moreover, on the boundaries of these regions we have $E_{ab}H^{ab} \equiv 0$,  and $H_{ab}$ satisfies $\fscalar{H} \equiv 0$ and $\ftensor{H}_{ab} \equiv0$. The physical interpretation of this result is then that no gravitational waves are allowed to pass through the boundaries of our lattice chambers, and so the lattice universe is piecewise silent. {\edit Such a result does not imply that gravitational waves play no part in the evolution of the space-time, but does illuminate their compatibility with the symmetries of the model.}

\section{Discussion}

In this paper we have studied the reflection symmetric surfaces that divide lattice universes into chambers. We have identified the restrictions that reflection symmetry imposes on kinematic and geometric quantities, and used the results to investigate the simplified equations that result for the expansion of these surfaces, as well as for the restrictions they impose on the propagation of gravitational waves.

We find that the area of the cell faces expands like the scale factor in a FLRW model that is filled with dust and radiation with energy flux, even though the space-time itself is completely devoid of any matter fields. We also find that the discrete symmetry of the configuration forces the scalar invariant $E^{ab}H_{ab}$ to vanish along all reflection symmetric surfaces, throughout the entire evolution of the model. Moreover, the only non-zero components of the magnetic Weyl tensor are shown to lie on the reflection surface itself. An explicit evaluation of the super-Poynting vector of the free gravitational field then {\edit demonstrates} that gravitational waves are forbidden from moving between chambers, {\edit as on the chamber boundaries there are not enough degrees of freedom in the space-time geometry to allow them to do so}. This shows that inhomogeneous discrete cosmological models with periodic boundary conditions are piecewise silent.

\ack

We are grateful to R Tavakol for helpful discussions. TC acknowledges support from the STFC. DG is supported by the Erasmus Mundus Joint Doctorate Program by Grant Number 2011-1640 from the EACEA of the European Commission.

\section*{Appendix. $1$+$2$-dimensional Covariant Equations for Gravity}

The twice contracted second Bianchi identities give the following conservation equations:
\begin{eqnarray}
\hspace{-20pt}\dot{\tilde{\rho}} + \theta (\tilde{\rho} + \tilde{p}) &=& - 2 \dot{u}^{\mu} \tilde{q}_{\mu} - D^{\mu} \tilde{q}_{\mu} - \varsigma_{\mu \nu} \tilde{\pi}^{\mu \nu}\\
\hspace{-20pt} \mathfrak{h}_{\mu}^{\phantom{\mu} \nu} \dot{\tilde{q}}_{\nu} +\frac{3}{2} \theta \tilde{q}_{\mu}  &=& -\varsigma_{\mu \nu} \tilde{q}^{\nu} - \varpi_{\mu \nu} \tilde{q}^{\nu} -(\tilde{\rho} +\tilde{p}) \dot{u}_{\mu} - D_{\mu} \tilde{p} - D^{\nu} \tilde{\pi}_{\mu \nu} - \tilde{\pi}_{\mu \nu} \dot{u}^{\nu} \, ,
\end {eqnarray}
where we have defined the projection tensor $\mathfrak{h}_{\mu \nu} := \gamma_{\mu \nu} + u_{\mu} u_{\nu}$. The Ricci identities give the following evolution equations:
\begin{eqnarray}
 \hspace{40pt} \dot{\theta}& =& -\frac{1}{2} \theta^2 - 2 \tilde{p} - 2 (\varsigma^2 - \varpi^2) +D^{\mu} \dot{u}_{\mu} + \dot{u}^{\mu} \dot{u}_{\mu} +2 \Lambda\\
\mathfrak{h}_{\mu}^{\phantom{\mu} \rho} \mathfrak{h}_{\nu}^{\phantom{\nu} \sigma} \dot{\varsigma}_{\rho \varsigma}& =& -\theta \varsigma_{\mu \nu} + D_{\langle \mu} \dot{u}_{\nu \rangle} + \dot{u}_{\langle \mu} \dot{u}_{\nu \rangle} + \tilde{\pi}_{\mu \nu}\\
\mathfrak{h}_{\mu}^{\phantom{\mu} \rho} \mathfrak{h}_{\nu}^{\phantom{\nu} \sigma} \dot{\varpi}_{\rho \sigma}& =& -\theta \varpi_{\mu \nu} +D_{[\nu} \dot{u}_{\mu]},
\end{eqnarray}
where $\varsigma^2 = \frac{1}{2} \varsigma_{\mu \nu} \varsigma^{\mu \nu}$ and $\varpi^2 = \frac{1}{2} \varpi_{\mu \nu} \varpi^{\mu \nu}$, and where we have included $\Lambda$. They also give the following constraint equation:
\beq
D^{\nu} \varsigma_{\mu \nu} - D^{\nu} \varpi_{\mu \nu} - 2 \varpi_{\mu \nu} \dot{u}^{\nu} - \frac{1}{2} D_{\mu} \theta + \tilde{q}_{\mu} =0,
\eeq
The scalar curvature of the 2-spaces orthogonal to $u^{\mu}$ is denoted ${}^{(2)}\mathcal{R}$, and can be used to write
\beq
\frac{\theta^2}{4} = \tilde{\rho} - \frac{{}^{(2)}\mathcal{R}}{2} + \varsigma^2 - \varpi^2 + \Lambda.
\eeq
Unlike the $1+3$-dimensional case, the uncontracted second Bianchi identities do not give any further equations beyond those stated above. For further details of the $1$+$2$-dimensional approach the reader is referred to \cite{doug}.


\section*{References}

\begin{thebibliography}{99}

\bibitem{planck} Planck Collaboration, arXiv:1303.5076 (2013).

\bibitem{perlmutter}
S. Perlmutter {\it et al.}, {\it Astrophys. J.} {\bf 517}, 565 (1999).

\bibitem{riess} 
A. Riess {\it et al.}, {\it Astron. J.} {\bf 116}, 1009 (1998).

\bibitem{Andersson&Coley:2011}
L.~Andersson, \& A.~Coley, {\it Class. Quant. Grav.} {\bf 28}, 160301 (2011).

\bibitem{perts}
C.~Clarkson \& O.~Umeh, {\it Class. Quant. Grav.} {\bf 28}, 164010 (2011).

\bibitem{space}
T.~Buchert \& S. R\"{a}s\"{a}nen, {\it Ann. Rev. Nuc. \& Part. Sci.} {\bf 62}, 57 (2012).

\bibitem{space-time}
T.~Clifton, A.~Coley \& R.~van~den~Hoogen, {\it JCAP} {\bf 10}, 044 (2012).

\bibitem{green}
S.~Green \& R.~Wald, {\it Phys. Rev. D} {\bf 83}, 084020 (2011).

\bibitem{bull}
P.~Bull \& T.~Clifton, {\it Phys. Rev. D} {\bf 85}, 103512 (2012).

\bibitem{larena}
J.-P.~Bruneton \& J.~Larena, {\it Class. Quant. Grav.} {\bf 29}, 155001 (2012).

\bibitem{Clifton_etal:2012}
T. Clifton, K. Rosquist \& R. Tavakol, {\it Phys. Rev. D} {\bf 86}, 043506 (2012).

\bibitem{Clifton_etal:2013}
T.~Clifton, D.~Gregoris, K.~Rosquist \& R.~Tavakol, {\it JCAP} {\bf 11}, 010 (2013).

\bibitem{mccrea}
W.~H.~M$^c$Crea, {\it Proc. Edinb. Math. Soc.} {\bf 2}, 158 (1931).

\bibitem{Coxeter&Whitrow:1950}
H.~S.~M.~Coxeter \& G.~J.~Whitrow, {\it Proc. Roy. Soc. A} {\bf 201}, 417 (1950).

\bibitem{Lindquist&Wheeler:1957}
R.~W.~Lindquist \& J.~A.~Wheeler, {\it Rev. Mod. Phys.} {\bf 29}, 432 (1957); {\it Rev. Mod.
Phys.} {\bf 31}, 839 (1959).

\bibitem{geocell}
J.~A.~Wheeler, {\it Found. Phys.} {\bf 13}, 161 (1983).

\bibitem{CF1}
T.~Clifton \& P.~G.~Ferreira, {\it Phys. Rev. D} {\bf 80}, 103503 (2009); {\it Phys. Rev. D} {\bf 84}, 109902 (2011).

\bibitem{CF2}
T.~Clifton \& P.~G.~Ferreira, {\it JCAP} {\bf 10}, 26 (2009).

\bibitem{CF3}
T.~Clifton, P.~G.~Ferreira \& K.~O'Donnell {\it Phys. Rev. D} {\bf 85}, 023502 (2012).

\bibitem{bentivegna1}
E. Bentivegna, M. Korzi\'{n}sky, {\it Class. Quant. Grav.} {\bf 29}, 165007 (2012).

\bibitem{bentivegna2}
E. Bentivegna, M. Korzi\'{n}sky, {\it Class. Quant. Grav.} {\bf 30}, 235008 (2013).

\bibitem{yoo1}
C. M. Yoo {\it et al.}, {\it Phys. Rev. D} {\bf 86}, 044027 (2012).

\bibitem{yoo2}
C. M. Yoo {\it et al.}, {\it Phys. Rev. Lett.} {\bf 111}, 161102 (2013).

\bibitem{jantzen}
R. T. Jantzen, P. Carini \& D. Bini, {\it Ann. Phys.} {\bf 215}, 1 (1992).

\bibitem{Maartens&Bassett:1998}
R. Maartens \& B. A. Bassett, {\it Class. Quant. Grav.} {\bf 15}, 705717 (1998).

\bibitem{coxeter}
H. M. S. Coxeter, ``Regular Polytopes'', Methuen and Company Ltd., London (1948).

\bibitem{mik}
M. Korzi\'{n}sky, arXiv:1312.0494 (2013).

\bibitem{bentivegna3}
E. Bentivegna, {\it Class. Quant. Grav.} {\bf 31}, 035004 (2013).

\bibitem{Borovik&Borovik:2010}
A.~V.~Borovik \& A.~Borovik, ``Mirrors and {R}eflections'', Springer (2010).

\bibitem{rindler}
W. Rindler, ``Relativity'', Oxford University Press (2001).

\bibitem{Chrusciel:1991}
P. T. Chru\'sciel, ``On Uniqueness in the Large of Solutions of Einstein's Equations (Strong Cosmic Censorship)", vol. 27 of Proceedings of the Centre for Mathematics and its Applications, Australian National University Press, Canberra, Australia (1991).

\bibitem{Friedrich&Rendall:2000}
H. Friedrich \& A. D. Rendall, in ``Einstein's Field Equations and Their Physical Implications: Selected Essays in Honour of J\"urgen Ehlers" B. G. Schmidt ed., vol. 540 of Lecture Notes in Physics, Springer (2000).

\bibitem{EllisLRS}
G.~F.~R.~Ellis, {\it J. Math. Phys.} {\bf 8}, 1171 (1966).

\bibitem{SKB}
G. F. R. Ellis \&  M. A. H. MacCallum, {\it Comm. Math. Phys.} {\bf 12}, 108 (1969).


\bibitem{uggla2}
H. van Elst \& C. Uggla, {\it Class. Quant. Grav.} {\bf 14}, 2673 (1997).

\bibitem{cosmological}
G. F. R. Ellis \& H. van Elst, \lq\lq Comological models\rq\rq, arXiv:gr-qc/9812046v5  (2008).

\bibitem{clarkson}
C. Clarkson, {\it Phys.Rev. D} {\bf 76}, 104034 (2007).

\bibitem{doug}
J.~D.~Barrow, D.~J.~Shaw \& C.~G.~Tsagas, {\it Class. Quant. Grav.} {\bf 23}, 5291 (2006).

\bibitem{ex}
J.~D.~Barrow \& T.~Clifton, {\it Phys. Rev. D} {\bf 73}, 103520 (2006).

\bibitem{Senovilla:2000}
J.~M.~M.~Senovilla, {\it Class. Quant. Grav.} {\bf 17}, 2799 (2000).

\bibitem{Zakharov:1973}
V.~D.~Zakharov, \lq\lq Gravitational waves in Einstein's theory\rq\rq, John Wiley \& sons, New York (1973).

\bibitem{Schutz:2009}
B. Schutz, \lq\lq A first course in general relativity\rq\rq, Cambridge University Press, Cambridge (2009).

\bibitem{Misner_etal:1973}
C.~W.~Misner, K.~S.~Thorne \& J.~A.~Wheeler, \lq\lq Gravitation\rq\rq, Freeman, New York (1973).

\bibitem{bonilla}
M.~\'{A}.~G. Bonilla \& J.~M.~M.~Senovilla, {\it Gen. Rel. Grav.} {\bf 29}, 91 (1997).

\bibitem{gomez}
A.~ Garc\'{i}a-Parrado G\'{o}mez-Lobo, {\it Class. Quant. Grav.} {\bf 25}, 015006 (2008).

\end{thebibliography}
\end{document}